\documentclass[12pt]{article}
\usepackage{graphicx}
\usepackage{amsmath}
\usepackage{amsfonts}
\usepackage{amsthm}
\usepackage{verbatim}

\begin{document}

\title{On the Mathematics of Fraternal Birth Order Effect and the Genetics of Homosexuality}
\author{Tanya Khovanova}

\date{September 5, 2017}

\maketitle

\begin{abstract}

Mathematicians have always been attracted to the field of genetics. I am especially interested in the mathematical aspects of research on homosexuality. Certain studies show that male homosexuality may have a genetic component that is correlated with female fertility. Other studies show the existence of the fraternal birth order effect, that is the correlation of homosexuality with the number of older brothers. 

This paper is devoted to the mathematical aspects of how these two phenomena are interconnected. In particular, I show that the fraternal birth order effect produces a correlation between homosexuality and maternal fecundity. Vice versa, I show that the correlation between homosexuality and female fecundity implies the increase of the probability of the younger brothers being homosexual.
\end{abstract}

Keywords: Fraternal birth order effect, male homosexuality, fecundity, genetics.


\section{Background}

According to the study by Blanchard and Bogaert \cite{BB} (1996): ``[E]ach additional older brother increased the odds of [male] homosexuality by 34$\%$." (see also Blanchard \cite{Blanchard} (2004) and Bogaert \cite{Bogaert} (2006) and a recent survey \cite{Blanchard2}). The current explanation is that carrying a boy to term changes their mother's uterine environment. Male fetuses produce H-Y antigens which may be responsible for this environmental change for future fetuses.

The research into a genetic component of male gayness shows that there might be some genes in the X chromosome that influence male homosexuality. It also shows that the same genes might be responsible for increased fertility in females (see \cite{CCZ} (2008) and \cite{IC} (2008)).

In this paper I compare two mathematical models. In these mathematical models I disregard girls for the sake of clarity and simplicity.

The first mathematical model of \textit{Fraternal Birth Order Effect}, which I denote MMFBOE, assumes that each next-born son becomes homosexual with increased probability. This probability is independent of any other factor.

The second mathematical model of \textit{Female Fecundity}, which I denote MMFF, assumes that a son becomes homosexual with probability depending on the total number of children and nothing else. 

I show mathematically how MMBOE implies correlation with family size and MMFF implies correlation with birth order. That means these two models are mathematically intertwined.

I also discuss the \textit{Brother Effect}. Brothers share a lot of the same genes. It is not surprising that brothers are more probable to share traits. With respect to homosexuality, I denote the correlation that homosexuals are more probable to have a homosexual brother than a non-homosexual as BE. The existence of genes that increase predisposition to homosexuality implies BE. The connection between MMFBOE and BE is more complicated.

I also discuss how to separate MMBOE and MMFF in the data.

Section~\ref{sec:ee} contains extreme mathematical examples that amplify the results of this paper. Section~\ref{sec:FBOEiFF} shows how MMFBOE implies the correlation with family size. Section~\ref{sec:FFiFBOE} shows how MMFF implies the correlation with birth order. In Section~\ref{sec:be} I discuss the connection between MMFBOE and the brother effect. In Section~\ref{sec:separate} I discuss how to separate the birth order from family size.

\section{Extreme Examples}\label{sec:ee}

First consider extreme theoretical examples. In the first two examples, suppose mothers only give birth to sons and only to one or two sons.

\textbf{First Extreme example.} This is an extreme variation of MMFBOE. Suppose the first son has a zero probability of being gay (which means that first sons are never gay) and the second son has probability one of being gay (which means he is always gay). Then all mothers of one son will have a straight son. All mothers with two sons will have one gay and one straight son. Homosexuals appear only in two-son families and never in one-son families. Therefore, MMFBOE implies the correlation with family size.

\textbf{Second Extreme example.} This is an extreme variation of MMFF. Suppose mothers with one son have probability zero of having a gay son. Suppose mothers with two sons have two homosexual sons with probability one. The first born is sometimes gay and sometimes straight, but the second son is always gay. Hence, it is more probable that the second son is gay. Therefore, MMFF implies the correlation with birth order.

These extreme variations of MMFBOE and MMFF show that these two models are intertwined. 

The next two sections explain this in more detail.

\section{MMFBOE and the family size}\label{sec:FBOEiFF}

Let us build a model with variables for numbers that correspond to MMFBOE. In this simple model we assume that the probability of a child being gay depends only on birth order and nothing else.

\textbf{MMFBOE model.} Let us assume that mothers have either one or two boys. Let $a$ be the probability of a woman having one boy, and correspondingly, $1-a$ of having two boys. Suppose $N$ is the total number of women in consideration. Suppose $p_1$ is the probability that the first boy is homosexual and $p_2$ is the probability that the second boy is homosexual. The fraternal birth order effect means that $p_2 > p_1$.

Now we produce the results of such a model.

Let us first estimate the total number of boys $T$:
$$T = aN + 2(1-a)N = (2-a)N.$$ 
The number of homosexuals in the one-son families is expected to be $ap_1N$. The expected number of homosexual first-born sons in two-son families is $(1-a)p_1N$ and the expected number of homosexual second-born sons is $(1-a)p_2N$. The total expected number of homosexuals $H$ is the sum:
$$H = p_1N + (1-a)p_2N.$$

The probability that a randomly chosen boy is a homosexual is 
$$\frac{H}{T} = \frac{p_1 + (1-a)p_2}{2-a}.$$

Let us see what happens with fecundity. Suppose we pick a mother randomly, then pick her son randomly. If there is only one son, then he is the one we have to pick. The probability that we pick a gay son, given that we picked the mother with one child is $p_1$. The probability that we pick a gay son, given that we picked the mother of two children is $(p_1+p_2)/2 > p_1$. This difference is the source of the correlation with fecundity.

To calculate this properly we need to choose a boy randomly and find the average fertility of the mother. The formula is given by the following equation:
\begin{equation}\label{eq:avfert}
\frac{\textit{\# number of single sons} + 2\cdot \textit{\# number of non-single sons}}{\textit{\# number of sons}}.
\end{equation}

First we calculate average maternal fertility per boy: 

For a randomly chosen boy (including both homosexual and non-homosexual boys), there are $aN$ mothers of one son and $(1-a)N$ mothers of two sons. Hence, a mother of a randomly chosen boy has on average 
\[\frac{aN + 2 \cdot 2(1-a)N}{(2-a)N}\]
children, which is equal to 
$$\frac{a +4-4a}{2-a} = 2 - \frac{a}{2-a}.$$

Let us see what happens with homosexual boys.  We have $ap_1N$ expected gay boys from one-son families and $(1-a)(p_1+p_2)N$ expected gay boys from two-son families. Now we plug this into the Eq.~(\ref{eq:avfert}) where we replace a randomly chosen boy with a gay boy to get:

\begin{eqnarray*}
& \frac{ap_1N + 2(1-a)(p_1+p_2)N}{ap_1N+(1-a)(p_1+p_2)N} &\\ 
& = \frac{ap_1 + 2(1-a)(p_1+p_2)}{ap_1 + (1-a)(p_1+p_2)} &\\
& = \frac{2(ap_1 + (1-a)(p_1+p_2))-ap_1}{ap_1 + p_1 - ap_1 +p_2 -ap_2} &\\
& = 2 - \frac{ap_1}{p_1+p_2-ap_2}. &
\end{eqnarray*}

If we denote by $c$ the ratio $p_2/p_1$, then the average maternal fertility per gay boy is
$$2 - \frac{a}{1 + (1-a)c}.$$ 
As $c > 1$, then $1 + (1-a)c > 2 -a$. Therefore, $2 - \frac{a}{1 + (1-a)c} > 2 - \frac{a}{2-a}$. It follows that the average maternal fertility per gay boy is greater than the overall average maternal fertility.

It is useful to note, that if $c=1$, then there is no correlation with the family size; that is, the average fertility is the same for randomly choen boys and gay boys. This is the expected result. Indeed, $c = 1$ means homosexuality does not depend on the birth order and is assigned completely randomly.

The impact of MMFBOE on the correlation with the family size is stronger if we consider larger families. Suppose $p_1 < p_2 < p_3 < \ldots$ are the probabilities of the first, second, and so on child being gay, correspondingly. Then the average probability, $x_k$, of being gay per child in a $k$-son family is 
$$x_k = \frac{p_1 + p_2 + \cdots + p_k}{k}.$$
When we add larger numbers to the average, the average increases. Thus, $x_j > x_i$, when $j > i$.

Among boys with many older brothers, there is a larger proportion of homosexuals. Thus, they contribute more to the calculation of average fecundity. Hence, if we add to our model the possibility of more than two boys where each next boy has a higher probability of being  homosexual, the correlation will be more impressive.

The results show that MMFBOE implies correlation with the family size. The female fecundity correlation with male homosexuality was shown not only for mothers, but also for maternal aunts and grandmothers \cite{IC}, \cite{CCZ}. This means, the fecundity results as a whole are not threatened by my examples. I will describe in Section~\ref{sec:separate} how to work with the data to mathematically separate birth order and family size.

\section{MMFF implies birth order correlation}\label{sec:FFiFBOE}

Let us build a mathematical model with variables instead of fixed numbers that correspond to the correlation of homosexuality with female fecundity. In this simple model we assume that the probability of a child being gay depends only on the family size and nothing else.

\textbf{MMFF model.} Let us assume that mothers have either one or two boys. Let $a$ be the probability of a woman having one boy, and correspondingly, $1-a$ of having two boys. Suppose $N$ is the total number of women in consideration. Suppose $q_1$ is the probability that a boy in a one-son family is homosexual and $q_2$ is the probability that a boy in a two-son family is homosexual. We assume that $q_2 > q_1$ to support the correlation of female fecundity with homosexuality. 

Here are the results of such a model. In our notation we use index $f$ for first sons and $s$ for second sons.

Let us see what happens with birth order. We start with first sons. The total number of first sons $T_f$ is $N$:
$$T_f = N.$$ 
The number of homosexuals in one-son families is expected to be $aq_1N$. The number of homosexual first sons in two-son families is expected to be $(1-a)q_2N$. The total number of first sons that are homosexual, $H_f$, is expected to be:
$$H_f = aq_1N + (1-a)q_2N.$$

The probability that the first-born is homosexual is
$$\frac{H_f}{T_f} = aq_1 + (1-a)q_2.$$

Now we do the same for the second-born sons. The expected total number of them $T_s$ is:
$$T_s = (1-a)N.$$ 
The expected number of homosexuals among them $H_s$ is:
$$H_s = (1-a)q_2N.$$

The probability that the second-born son is homosexual is
$$\frac{H_s}{T_s} = q_2.$$

The final mathematical step needs to show that the probability that the first born is homosexual is less than the probability that the second born is homosexual. It follows from the fact that $q_1 < q_2$. Indeed:
$$\frac{H_f}{T_f} = aq_1 + (1-a)q_2 < aq_2 + (1-a)q_2 = q_2 = \frac{H_s}{T_s}.$$

If we denote by $c$ the ratio $q_2/q_1$, then the ratio of increase, that is $\frac{H_s}{T_s}$ divided by $\frac{H_f}{T_f}$ is:
$$\frac{q_2}{aq_1 + (1-a)q_2} = \frac{c}{a + (1-a)c}.$$ 

It is useful to note, that if $c=1$, then there is no correlation with the birth order: the first-born sons and second-born sons are homosexuals with the same probability.

The impact of MMFF on the birth order is stronger if we consider larger families. Suppose $q_1 < q_2 < q_3 < \ldots$ are the probabilities of sons being homosexual in families of size 1, 2, and so on, respectively. Then the average probability $y_i$ of a child number $i$ being gay depends on the distribution of family sizes. Suppose the number of families of size $m$ is $N_m$, then we can calculate $y_i$ as:
$$y_i = \frac{q_iN_i + q_{i+1}N_{i+1} + q_{i+2}N_{i+2} + \cdots}{N_i + N_{i+1} + N_{i+2} +\cdots}.$$

We can show that $y_j > y_i$, when $j> i$. Let us denote $Q_1 = q_iN_i + q_{i+1}N_{i+1} + q_{i+2}N_{i+2} + \cdots + q_{j-1}N_{j-1}$ and $Q_2 = q_jN_j + q_{ij+1}N_{j+1} + q_{j+2}N_{j+2} + \cdots$. Further, let us denote $M_1 = N_i + N_{i+1} + N_{i+2} +\cdots + N_{j-1}$ and $M_2 = N_j + N_{j+1} + N_{j+2} +\cdots$. Then 
$y_i = \frac{Q_1 + Q_2}{M_1 + M_2}$
and
$$y_j = \frac{Q_2}{M_2}.$$
The important observation is that
$$\frac{Q_1}{M_1} \leq q_{i-1} < q_i < \frac{Q_2}{M_2}.$$
Therefore,
$$Q_1M_2 < Q_2M_1.$$
It follows that $$Q_1M_2 + Q_2M_2 < Q_2M_1+ Q_2M_2.$$
This implies
$$y_i = \frac{Q_1 + Q_2}{M_1 + M_2} < \frac{Q_2}{M_2} = y_j.$$

The results show that MMFF implies correlation with the birth order. The fraternal birth order effect was shown only for brothers and not for sisters. This means, the FBOE as a whole is not threatened by my examples. I will describe in Section~\ref{sec:separate} how to separate birth order and family size mathematically in the data.

\section{Brothers}\label{sec:be}

Siblings share a lot of genetic material. Not surprisingly they have a lot of common traits. If a trait is genetic, then the probability that a sibling has it is higher than the probability that a randomly chosen person has it. Very often the fact that siblings share traits with higher probability than random people share traits serves as a confirmation that the trait is genetic. That means the existence of a homosexual gene would imply the higher probability that a gay person has a gay brother than the probability that a randomly chosen person has a gay brother. I denoted this correlation BE.

Is there a mathematical way to connect the birth order with BE? The answer: it is complicated. 

Let us look at how the fraternal birth order effect influences the probability that a gay boy has a gay brother. The probability that a gay person has a gay brother depends on the number of boys in the family. If a boy does not have brothers he cannot have a gay brother. If a boy has a million brothers, then with extremely high probability at least one of them will be gay.

Here are two extreme mathematical examples where we assume that mothers have only one or three sons.

\textbf{Third extreme example.} This is an extreme variation of MMFBOE. Suppose the first son has zero probability of being gay and the second and third sons have probability one of being gay.  That is, $p_1=0$ and $p_2 = p_3 = 1$. All gay boys in this model have a gay brother, while a randomly chosen boy sometimes has one and sometimes does not. With these particular probabilities MMFBOE implies BE.

\textbf{Fourth extreme example.} This is an extreme variation of MMFBOE. Suppose the first and second sons have zero probability of being gay and the third son has probability one of being gay. That is, $p_1=p_2 = 0$ and $p_3 = 1$. No gay boy in this model has a gay brother, while some randomly chosen boys have one. With these particular probabilities MMFBOE contradicts BE.

It follows that depending on the actual numbers MMBOE might or might not imply BE.

\section{Separating birth order and female fecundity}\label{sec:separate}

Our simplistic models in Sections~\ref{sec:FBOEiFF} and~\ref{sec:FFiFBOE} showed that MMFBOE and MMFF imply each other. That means birth order and female fecundity are intertwined in the data. It is important to separate these two different models.  To do it we need to fix some variables.

\textbf{Method 1.} To show how the birth order works independently of female fecundity, we need to fix the family size. Suppose we consider only families of size 2. Then the fertility does not play a role. In this case, according to the fraternal birth order effect, the second son is gay with higher probability than the first son. The corresponding probabilities derived from real data should confirm MMFBOE without interference of MMFF.

\textbf{Method 2.} To show how the female fecundity works independently of the fraternal birth order effect, we need to consider only the first sons. Then the MMFBOE does not play a role. In this case, according to MMFF, the first son in a larger family is gay with higher probability than the first son in a smaller family. The corresponding probabilities derived from real data should confirm MMFF without MMFBOE.

Consider the theoretical discussion of families of size one and two in Sections~\ref{sec:FBOEiFF} and~\ref{sec:FFiFBOE}. Here is the joint mathematical model, which I call MMFBOE-FF.

\textbf{MMFBOE-FF}. Let us consider only the case of women who have one or two boys. Let $a$ be the probability of a woman having one boy, and correspondingly, $1-a$ of having two boys. Suppose $N$ is the total number of women in consideration. Suppose $p_{11}$ is the probability that the first boy in a one-son family is homosexual, $p_{12}$ is the probability that the first boy in a two-son family is homosexual, and $p_{22}$ is the probability that the second boy in a two-son family is homosexual. The female fecundity means $p_{12}>p_{11}$. The fraternal birth order effect means $p_{12} > p_{22}$.


\begin{thebibliography}{9}

\bibitem{Blanchard} Blanchard, R. (2004). Quantitative and theoretical analyses of the relation between older brothers and homosexuality in men. \textit{Journal of Theoretical Biology} \textbf{230}  173--187.

\bibitem{Blanchard2} Blanchard, R. (2017). Fraternal Birth Oder, Family Size, and Male Homosexuality: Meta-Analysis of Studies Spanning 25 Years. \textit{Arch. Sex. Behav.}

\bibitem{BB} Blanchard, R., \& Bogaert, A.~F. (1996). Homosexuality in Men and Number of Older Brothers. \textit{The American Journal of Psychiatry}  \textbf{153}, 1, 27--31.


\bibitem{Bogaert} Bogaert A.~F. (2006). Biological versus nonbiological older brothers and men's sexual orientation. \textit{PNAS}, \textbf{103}, 28, 10771--10774.

\bibitem{CCZ} Ciani, A.~C., Cermelli, P. \& Zanzotto, G. (2008). Sexual\-ly Antagonistic Selection in Human Male Homosexuality. \textit{PLoS ONE}, \emph{http://\-www.plosone.org/\-article/\-info:\-doi\%2F10.\-1371\%2F\-jour\-nal.\-pone.0002282}.

\bibitem{IC} Iemmola, F. \& Ciani, A.~C. (2008). New Evidence of Genetic Factors Influencing Sexual Orientation in Men: Female Fecundity Increase in the Maternal Line. \textit{Arch Sex Behav}.

\bibitem{Kh} Khovanova, T. (2008) Fraternal Birth Order Threatens Research into the Genetics of Homosexuality, Tanya Khovanova's Math Blog, \emph{http://\-blog.tanya\-khovanova.com/?p=40}.






\end{thebibliography}
\end{document}